\documentclass[sigconf]{acmart}
\acmConference[MSR 2024]{21st International Conference on Mining Software Repositories}{April 2024}{Lisbon, Portugal}
\usepackage[linesnumbered,ruled]{algorithm2e}
\usepackage{graphicx}

\newif\ifSPACEHACK
\SPACEHACKtrue 

\newif\ifDEBUG
\DEBUGfalse

\newif\ifANONYMOUS
\ANONYMOUSfalse

\newif\ifARXIV
\ARXIVfalse

\usepackage{amsmath,amssymb,amsfonts}

\usepackage{algorithmic}
\usepackage{graphicx}
\usepackage{textcomp}
\usepackage{xcolor}
\usepackage{booktabs} 
\usepackage{xspace} 
\usepackage[normalem]{ulem}
\usepackage{makecell}
\usepackage{tcolorbox}
\usepackage{enumitem}
\usepackage{siunitx}
\usepackage[vskip=1em,font=itshape,leftmargin=2em,rightmargin=2em]{quoting}
\usepackage{tabularx}
\usepackage{hyperref}
\usepackage{lscape}
\usepackage{afterpage}

\usepackage{makecell}
\usepackage{multirow}
\usepackage{booktabs}
\usepackage[T1]{fontenc}

\usepackage{setspace}

\usepackage{tikz}

\usepackage{wrapfig}

\usepackage{lipsum}
\usepackage[noadjust]{cite}

\usepackage{soul}

\ifDEBUG
    \newcommand{\JD}[1]{\textcolor{purple}{[JD:#1]}}
    \newcommand{\AG}[1]{\textcolor{olive}{[AG:#1]}}
    \newcommand{\WJ}[1]{\textcolor{teal}{[WJ:#1]}}
    \newcommand{\GKT}[1]{\textcolor{brown}{[GKT:#1]}}
    
    \newcommand{\NV}[1]{\textcolor{red}{[NV: #1]}}
    \newcommand{\CC}[1]{\textcolor{violet}{[CC: #1]}}
    \newcommand{\JY}[1]{\textcolor{cyan}{[JY: #1]}}
     \newcommand{\YT}[1]{\textcolor{red}{[YT: #1]}}

    \newcommand{\TODO}[1]{\hl{#1}}
\else
    \newcommand{\JD}[1]{}
    \newcommand{\AG}[1]{}
    \newcommand{\WJ}[1]{}
    \newcommand{\GKT}[1]{}
    \newcommand{\NS}[1]{}
    \newcommand{\NV}[1]{}
    \newcommand{\CC}[1]{}
    \newcommand{\JY}[1]{}
    \newcommand{\YT}[1]{}

    \newcommand{\TODO}[1]{#1}
\fi

\usepackage[compact]{titlesec}
    \usepackage{etoolbox}
    \makeatletter
    \patchcmd{\ttlh@hang}{\parindent\z@}{\parindent\z@\leavevmode}{}{}
    \patchcmd{\ttlh@hang}{\noindent}{}{}{}
    \makeatother

\newcommand{\myparagraph}[1]{\vspace{0.10cm}\noindent\textbf{#1} \noindent{}}

\ifSPACEHACK
    \titlespacing*\section{0pt}{5pt plus 1pt minus 1pt}{3pt plus 1pt minus 1pt}
    \titlespacing*\subsection{0pt}{4pt plus 1.5pt minus 1.5pt}{4pt plus 1.5pt minus 1.5pt}
    \titlespacing*\subsubsection{0pt}{3pt plus 1pt minus 1pt}{3pt plus 1.5pt minus 1.5pt}
    \titlespacing*\paragraph{0pt}{1pt plus 1.5pt minus 1.5pt}{2pt plus 1.5pt minus 1.5pt}
\fi

\usepackage{balance} 
\usepackage[font=small,labelfont=bf]{caption}
\usepackage{subcaption}
\usepackage[bottom]{footmisc}
\makeatletter
\def\cl@chapter{}
\makeatother
\usepackage{cleveref}

\crefformat{section}{\S#2#1#3}
\crefname{figure}{Figure}{Figures}
\crefname{appendix}{Appendix}{Appendices}
\crefname{table}{Table}{Tables}
\crefname{algorithm}{Algorithm}{Algorithms}
\crefname{listing}{Listing}{Listings}
\crefname{theorem}{Theorem}{Theorems}
\crefname{thm}{Theorem}{Theorems}
\crefname{lemma}{Lemma}{Lemmata}
\crefname{equation}{Eqt.}{Eqts.}
\crefformat{Grammar}{Grammar #1}

\usepackage{enumitem}
\usepackage{booktabs}

\newcommand{\ie}{\textit{i.e.,} }
\newcommand{\eg}{\textit{e.g.,} }
\newcommand{\etal}{\textit{et al.}\xspace}

\newcommand{\code}[1]{{\small\texttt{#1}}\xspace}

\newcommand{\DatasetNickname}{\emph{PeaTMOSS}\xspace}

\newcommand{\DatasetNicknameFormatted}{\code{\textbf{P}ea\textbf{TMOSS}}\xspace}
\newcommand{\DatasetNicknameExpanded}{\textbf{\ul{P}}re-\textbf{\ul{T}}rained \textbf{\ul{M}}odels in \textbf{\ul{O}}pen-
\textbf{\ul{S}}ource \textbf{\ul{S}}oftware\xspace}

\newcommand{\PercentageOfAppNoLicense}{{{43.42\%}}\xspace}
\newcommand{\PercentageOfIdenticalLicense}{{{25.61\%}}\xspace}
\newcommand{\PercentageOfUnknownLicensePairs}{{{47.33\%}}\xspace}
\newcommand{\PercentageOfInconsistentLicensePairs}{{{0.24\%}}\xspace}

\newcommand{\TotalNumberOfLinks}{{44,337}\xspace}
\newcommand{\NumberOfPTPTMs}{{{362}}\xspace} 
\newcommand{\NumberOfHFPTMs}{{{281,276}}\xspace} 
\newcommand{\PTMPTMDepencencyCount}{{8,829}\xspace} 
\newcommand{\NEWNEWnumberOfPTMRepos}{{14,296}\xspace} 
\newcommand{\GitHubReuseRepoCount}{{15,129}\xspace}

\newcommand{\NumberOfReusedPTMs}{{2,530}\xspace} 

\newcommand{\GPTThreeEval}{{67.46\%}\xspace}
\newcommand{\GPTFourEval}{{94.39\%}\xspace}

\newcommand{\NEWnumberOfPTMs}{{281,638}\xspace}

\newcommand{\NEWDatasetSize}{{48.2 TB}\xspace}
\newcommand{\NEWSizeOfMetadata}{{7.12 GB}\xspace}
\newcommand{\NEWSizeOfRepos}{\TODO{xxx TB}\xspace}

\newcommand{\DBFileName}{\code{PeaTMOSS.db}\xspace}

\newcommand{\GitHubReuseLibrariesTotal}{{27}\xspace}
\newcommand{\GitHubReuseSignatureTotal}{{474}\xspace}

\newcommand{\GitHubReuseRepoCountSourceGraph}{{36,888}\xspace}
\newcommand{\GitHubReuseRepoCountTP}{{28,575}\xspace}
\newcommand{\GitHubReuseRepoCountStaticMatch}{{15,129}\xspace}

\newcommand{\GitHubReuseRepoCountSourceGraphSize}{{3.5 TB}\xspace}

\newcommand{\GitHubReusePrCount}{{12,184}\xspace}

\newcommand{\GitHubReuseIssueCount}{{19,513}\xspace}

\newcommand{\GitHubReuseRepoSourceGraphAvgStar}{{201}\xspace}
\newcommand{\GitHubReuseRepoSourceGraphDate}{{July 10, 2023}\xspace}

\newcommand{\GitHubReuseRepoCountSourceGraphHFFive}{{19,322}\xspace}

\newcommand{\TotalNumberOfPackagesMetadata}{{\NEWnumberOfPTMs}\xspace}

\begin{document}




\title{Extracting Pre-trained Model Metadata Using Large Language Model} 
\title{Investigating Pre-trained Deep Learning Model Adoption in Software Repositories
}
\title{PeaTMOSS: \ul{M}ining \ul{P}re-\ul{T}rained \ul{M}odels in \ul{O}pen-\ul{S}ource \ul{S}oftware}
\title{PeaTMOSS: A Dataset and Initial Analysis of \\ \ul{P}re-\ul{T}rained \ul{M}odels in \ul{O}pen-\ul{S}ource \ul{S}oftware}

\ifANONYMOUS
  \author{Anonymous Author(s)}
\else
    \author{Wenxin Jiang}
    \orcid{0000-0003-2608-8576}
    \affiliation{%
      \institution{Purdue University}
      \country{W Lafayette, IN, USA}}
    \email{jiang784@purdue.edu}

    \author{Jerin Yasmin}
    \orcid{0009-0001-9831-714X}
    \affiliation{%
      \institution{Queen's University}
      \country{Kingston, ON, CA}}
    \email{19jy2@queensu.ca}

    \author{Jason Jones}
    \orcid{0009-0005-7088-0597}
    \affiliation{%
      \institution{Purdue University}
      \country{W Lafayette, IN, USA}}
    \email{jone2078@purdue.edu}

     \author{Nicholas Synovic}
     \orcid{0000-0003-0413-4594}
     \affiliation{%
        \institution{Loyola University Chicago}
        \country{Chicago, IL, USA}
        }
    \email{nsynovic@luc.edu}

    \author{Jiashen Kuo}
    \orcid{0009-0000-1431-2249}
    \affiliation{%
      \institution{Purdue University}
      \country{W Lafayette, IN, USA}}
    \email{kuo90@purdue.edu}

    \author{Nathaniel Bielanski}
    \orcid{0009-0001-1453-9168}
    \affiliation{%
      \institution{Purdue University}
      \country{W Lafayette, IN, USA}}
    \email{nbielans@purdue.edu}

    \author{Yuan Tian}
    \orcid{0000-0002-2208-3893}
    \affiliation{%
      \institution{Queen's University}
      \country{Kingston, ON, CA}}
    \email{y.tian@queensu.ca}

    \author{George K. Thiruvathukal}
    \orcid{0000-0002-0452-5571}
    \affiliation{%
      \institution{Loyola University Chicago}
    \country{Chicago, IL, USA}
    }
    \email{gkt@cs.luc.edu}
    
    \author{James C. Davis}
    \orcid{0000-0003-2495-686X}
    \affiliation{%
      \institution{Purdue University}
      \country{W Lafayette, IN, USA}}
    \email{davisjam@purdue.edu}
\fi


\begin{abstract}
The development and training of deep learning models have become increasingly costly and complex. Consequently, software engineers are adopting pre-trained models (PTMs) for their downstream applications.
The dynamics of the PTM supply chain remain largely unexplored, signaling a clear need for structured datasets that document not only the metadata but also the subsequent applications of these models.
Without such data, the MSR community cannot comprehensively understand the impact of PTM adoption and reuse.


This paper presents the \DatasetNickname dataset, which comprises metadata for \TotalNumberOfPackagesMetadata PTMs and detailed snapshots for all PTMs with over 50 monthly downloads (\NEWNEWnumberOfPTMRepos PTMs), along with \GitHubReuseRepoCountTP open-source software repositories from GitHub that utilize these models.
Additionally, the dataset includes \TotalNumberOfLinks mappings from \GitHubReuseRepoCount downstream GitHub repositories to the \NumberOfReusedPTMs PTMs they use. 
To enhance the dataset's comprehensiveness, we developed prompts for a large language model to automatically extract model metadata, including the model's training datasets, parameters, and evaluation metrics.
Our analysis of this dataset provides the first summary statistics for the PTM supply chain, showing the trend of PTM development and common shortcomings of PTM package documentation. 
Our example application reveals inconsistencies in software licenses across PTMs and their dependent projects.
PeaTMOSS lays the foundation for future research, offering rich opportunities to investigate the PTM supply chain. 
We outline mining opportunities on PTMs, their downstream usage, and cross-cutting questions.

Our artifact is available at {\small \url{https://github.com/PurdueDualityLab/PeaTMOSS-Artifact}}.
Our dataset is available at {\small \url{https://transfer.rcac.purdue.edu/file-manager?origin\_id=ff978999-16c2-4b50-ac7a-947ffdc3eb1d\&origin\_path=\%2F}}.
\end{abstract}

\keywords{Datasets, Machine learning, Deep neural networks, Model zoos, Package registries, Open-Source, Empirical Software Engineering}

\maketitle
\renewcommand{\shortauthors}{Wenxin Jiang et al.}

\section{Introduction} \label{sec:intro}

\begin{figure}[ht]
    \centering
    \includegraphics[width=0.98\linewidth]{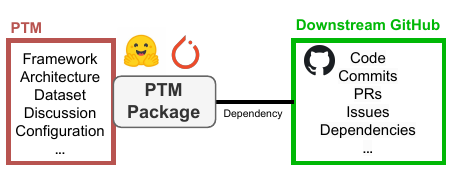}
    \caption{
    This paper presents the \DatasetNickname dataset: \DatasetNicknameExpanded.
    \DatasetNickname includes data on
      \NEWnumberOfPTMs pre-trained models,
      \GitHubReuseRepoCountTP GitHub repositories that use pre-trained models,
      and 
      \TotalNumberOfLinks links between them.
    }
    \label{fig:Overview}
\end{figure}

Deep Neural Networks (DNNs) have become 
a common component in software systems over the past decade. 
Developing and training DNN models is costly, requiring specialized hardware and large datasets~\citep{patterson2021carbon, geiping2023TrainingLMonSingleGPUinOneDay}.
While some software engineers develop DNNs from scratch, others integrate DNNs into software following a typical reuse pattern~\citep{marcelino_transfer_2022, kalyan2021SurveyofTransformerBasedPTMinNLP, davis2023JVA}:
  (1) pre-trained DNN models (PTMs) are published to registries such as Hugging Face (analogous to traditional package registries such as NPM);
  and 
  (2) other software depends on these PTMs, accessed by library or web API.

Despite the widespread adoption of PTMs~\citep{kathikar2023assessingHFVulnerabilities, shen2023hugginggpt}, 
our understanding of the software engineering practices and challenges surrounding PTM reuse remains limited~\citep{Jiang2022PTMReuse}. This understanding is critical for developing more sophisticated tools, mitigating risks, and guiding best practices~\citep{Amershi2019SE4MLCaseStudy}.
Mining Software Repositories techniques could help, but unfortunately current datasets on PTMs
lack crucial details, which leaves a gap in knowledge~\citep{jiang2023ptmtorrent, ait2023hfcommunity}. For instance, they frequently omit comprehensive evaluation metrics, model training conditions, parameters, and standardization in reporting results.
This absence of information impedes our ability to perform robust analyses, compare performances meaningfully, or derive a coherent picture of PTMs' impact and usage in software engineering. 
Recent work highlights the need for a more complete spectrum of metadata is required~\citep{Jiang2022PTMReuse, jiang2023ptmtorrent}, which should include---but not be limited to---details on model training datasets, versioning, licensing, and the computational requirements for PTM reuse.

To address this gap, the primary contribution of this work is the creation of the \DatasetNicknameFormatted dataset: \DatasetNicknameExpanded.
\DatasetNickname enables mining of PTMs, the software projects that use them, and the interactions between PTMs and downstream use. 
As illustrated in~\cref{fig:Overview},
\DatasetNickname contains a snapshot of: 
    (1) \NEWnumberOfPTMs PTMs,
    (2) \GitHubReuseRepoCountTP open-source software repositories that use PTMs, providing real-world context for how these models are applied, 
    and
    (3) \TotalNumberOfLinks mappings between PTMs and downstream GitHub repositories.
Our secondary contribution involves the practical application of large language models (LLMs) to extract PTM metadata, thereby enhancing our dataset (\cref{sec:enhanced-peatmoss}). 
We apply this tool to systematically extract key metadata, including datasets, hyper-parameters, and performance metrics, from unstructured model cards.
Li \etal called for comprehensive metadata to construct a queryable model zoo, enabling efficient search and comparison of models~\citep{Li2022MetadataRepresentation4QueryableMLModelZoos}.
By addressing the challenges of
unstructured data, we ensure that our model zoo encompasses a wide range of critical information, facilitating more informed and precise queries.

We conduct two demonstrations of the value of this dataset. 
    In~\cref{sec:DataAnalysis} we analyze the data distribution in \DatasetNickname, show the trends in the growth of PTM development and identify the common shortcomings in PTM package documentation.
    In~\cref{sec:ExampleApplication} we use the mapping created for PTM and GitHub projects to analyze the consistency of software licenses.
As future work, the PeaTMOSS dataset offers many opportunities to study and inform our understanding of the PTM supply chain.
We propose three distinct directions for analyzing \DatasetNickname:
    (1) analyses focusing on the GitHub data subset,
    (2) explorations centered on the PTM aspect,
    and (3) comprehensive studies integrating insights from both GitHub and PTM components.
We suggest researchers take advantage of the \DatasetNickname dataset and conduct a larger-scale measurement on characterizing the properties of the PTM supply chain.
\textbf{Our contributions are}:

\begin{itemize}[leftmargin=0.5cm]
    \item We share a dataset named PeaTMOSS which includes \TotalNumberOfPackagesMetadata PTM packages, and
    \GitHubReuseRepoCountTP downstream GitHub repositories.
    \item We tackled the issue of unstructured attributes by developing a LLM-based tool for metadata extraction, which enhances our dataset by adding structured data in JSON format.
    \item We provide the first summary statistics of this PTM supply chain, encompassing distributions of PTMs and their downstream repositories across various problem domains. Our analysis also includes trends in model size and the quantity of PTM packages, along with an overview of the proportion of available metadata.
    We show the proportion of missing data in each PTM metadata category. 
    \item We applied our dataset to assess the compatibility of PTMs with downstream GitHub repositories.
    Our findings reveal that \PercentageOfInconsistentLicensePairs of these licenses are inconsistent, potentially causing community confusion and hindering collaboration.
\end{itemize}

\vspace{0.1cm}
\noindent
\ul{Significance:}
\DatasetNickname is a comprehensive dataset for PTM in open-source software. It offers an extensive mapping between PTM packages and downstream GitHub repositories, and many queryable metadata.
Using \DatasetNickname, researchers can study the PTM supply chain and the reuse modes of PTM packages.
Engineering tools can be developed for PTM reuse, \eg for model search and comparison. 

\vspace{0.1cm}
\noindent
\ul{Paper outline:}
This paper is organized as follows:
  \cref{sec:background} and \cref{sec:RelatedWork} provide background and related work. 
In \cref{sec:original-Peatmoss}, we describe the original version of the \DatasetNickname dataset, and \cref{sec:enhanced-peatmoss} details the augmented dataset enriched with our metadata extraction pipeline.
Data analysis of \DatasetNickname is presented in \cref{sec:DataAnalysis}.
\cref{sec:ExampleApplication} illustrates a practical application.
The paper concludes with an examination of potential threats to validity in \cref{sec:threats}, followed by a discussion of future work in \cref{sec:FutureWork}.


\section{Background} \label{sec:background}

This section covers 
  PTMs (\cref{sec:background-PTM})
  and 
  their reuse (\cref{sec:background-PTMSupplyChain}).

\subsection{Pre-Trained Deep Learning Models (PTMs)} 
\label{sec:background-PTM}

The advent of deep learning has precipitated a fundamental shift in computational methodologies, transitioning from the deterministic algorithms characteristic of traditional software to increasingly probabilistic and data-driven paradigms~\citep{karpathy_software_2017}. Deep learning typically operates through neural networks capable of assimilating datasets, thereby enabling them to make predictions or perform complex tasks~\citep{lecun_deep_2015}. 
A PTM embodies a DNN architecture that has undergone prior training with a specific dataset, incorporating a defined data pipeline, training regime, and learned parameters (``weights'').
This pre-training equips the PTM to perform inference or to be adapted for downstream applications~\citep{davis2023JVA}.

Existing research has explored various methods for reusing deep learning models, such as feature extraction, transfer learning, data generation, and model compression~\citep{Han2021PTM, Jiang2022PTMSupplyChain}.
For instance, DNNs can be pre-trained using large-scale unlabeled molecular databases and then fine-tuned over
specific chemical downstream tasks like molecular property prediction~\citep{xia2023systematicSurveyofChemicalPTM}.
Additionally, models can be employed to annotate data or to synthesize new datasets through generative approaches~\citep{Dube2019DataLabeling4TransferLearning, antoniou2017dataAugGenerativeAdvNetworks}.
Transfer learning enables models trained on generic datasets to refine their understanding of more detailed, downstream tasks, often resulting in enhanced performance on specialized datasets~\citep{transferlearning}.
Furthermore, models can be optimized for size and efficiency to run on edge devices, a process known as model compression~\citep{Deng2020ModelCompressionandHWAccelerationforNN}. 

Thanks to this range of reuse modes, in recent years PTMs have become increasingly popular~\citep{kathikar2023assessingHFVulnerabilities, castano2023exploring}. 
The total number of open-source PTM packages has seen a consistent increase on a monthly basis~\citep{castano2023exploring}.
\cref{fig:PackagePopularity} provides a quantitative demonstration of the extensive adoption and rising popularity of PTMs. 
Previous research indicates that the popularity and adoption rate of Hugging Face's models are comparable to those of other established software package registries, including npm and PyPI~\citep{Jiang2022PTMReuse}.

\begin{table}[h]
\centering
\caption{
  Comparison of package counts and download figures for the top 10\% of PTMs on Hugging Face.
  Data for August 2022 is sourced from the PTMTorrent dataset~\citep{jiang2023ptmtorrent}.
  The August 2023 data is obtained from our dataset.
  This comparison highlights the growth of PTM usage over a one-year period (\ie doubling).
  }
\label{tab:popularity-comparison}
\begin{tabular}{lcc}
\toprule
\textbf{Hugging Face Statistics} & \textbf{Aug. 2022} & \textbf{Aug. 2023} \\
\midrule
\# packages of all PTMs & 124\ K  & 282\ K \\
\# downloads of top 10\% PTMs & 269 billion & 587 billion \\
\bottomrule
\label{fig:PackagePopularity}
\end{tabular}
\end{table}





\subsection{Components of the PTM Supply Chain} \label{sec:background-PTMSupplyChain}

\begin{figure}
    \centering
    \includegraphics[width=0.98\linewidth]{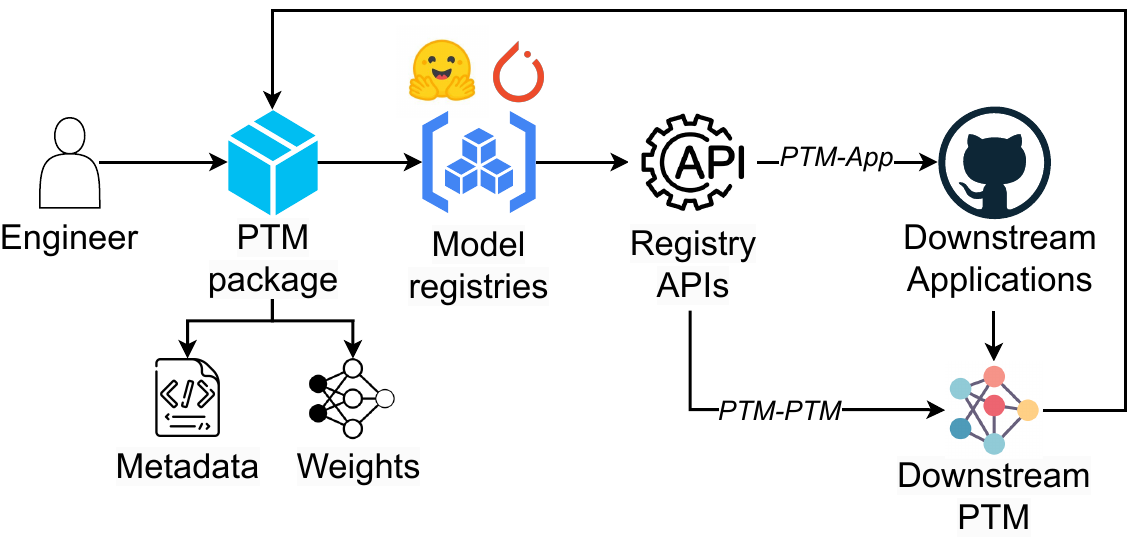}
    \caption{
    The PTM supply chain.
    Engineers publish PTM packages to model registries. 
    PTMs are used by applications and other PTMs.
    }
    \label{fig:PTMSupplyChain}
\end{figure}

Jiang \etal. introduced the PTM supply chain concept, encompassing PTM packages, the model registries, the authors of PTMs, and the users~\citep{Jiang2022PTMSupplyChain}.
\cref{fig:PTMSupplyChain} extends their model to include the downstream applications, thus providing a holistic view of the PTM ecosystem.
\DatasetNickname contains the major elements of this supply chain.
This section describes each element in turn.



\subsubsection{PTM Packages.}
PTMs are often shared in \textit{PTM packages}.
Per Jiang \etal~\citep{Jiang2022PTMReuse},
  a PTM package is analogous to traditional software packages on platforms like NPM or PyPI~\citep{npmWeb, pypiWeb}.
A PTM package has standard elements such as a license, documentation, and usage examples.
Analogous to source code, a PTM package describes the model architecture and pre-trained weights.
Its metadata indicates the training regime, which
  includes the dataset(s) involved,
  how the model's parameters were initialized,
  and
  the necessary data pre- and post-processing (``data pipeline'').
A PTM package may indicate the model's performance on evaluation metrics.

\subsubsection{Model Registries.}

PTM packages are commonly disseminated via deep learning model registries (also known as model hubs/zoos).
Jiang \etal define a deep learning model registry as \textit{a collaborative hub where teams share deep learning models}~\citep{Jiang2022PTMReuse}.
Prior work shows that there are three kinds of model registries categorized by their contribution types~\citep{Jiang2022PTMSupplyChain}:
  open (\eg Hugging Face~\citep{HuggingFaceWeb}),
  gated (\eg PyTorch Hub~\citep{PytorchHub}),
  and commercial (\eg NVIDIA NGC catalog~\citep{NVIDIANGC}).
These platforms enable engineers to directly adopt PTMs or adapt them through fine-tuning for specialized downstream tasks.

\subsubsection{Package Dependencies.}

The various methodologies for PTM reuse establish two distinct types of dependencies within the PTM supply chain.
Firstly, there are \textit{PTM-PTM} dependencies, where, for instance, one model might be fine-tuned from another~\citep{Jiang2022PTMSupplyChain}.
Secondly, there are \textit{PTM-Application} dependencies, where software projects rely on PTMs for their functionality~\citep{kathikar2023assessingHFVulnerabilities}.
These dependencies underscore the interconnected nature of PTM reuse and highlight an aspect of PTM package usage that necessitates further exploration, particularly in how these dependencies impact the broader software engineering landscape.

\section{Related Work} \label{sec:RelatedWork}

This section covers related work on software engineering in
  PTM reuse (\cref{sec:SEinPTMReuse}),
  importance of queryable PTM metadata (\cref{sec:ImportanceofQueryableMetadata}),
  and
  open-source PTM datasets (\cref{sec:background-OOSDataset}).

\subsection{Software Engineering in PTM Reuse}
\label{sec:SEinPTMReuse}
Prior work has comprehensively studied the development of deep learning systems from software engineering perspectives~\citep{Amershi2019SE4MLCaseStudy, Rahman2019MLSEinPractice}.
These works more focused on creating and training new DNNs from scratch, which usually requires extensive resources and expertise. 
However, the reuse process of PTM focused on adapting existing PTMs which is a different process compared to developing a new model~\citep{Jiang2022PTMReuse}.
The literature of understanding the reuse of PTMs still presents a notable gap.

Davis \etal introduced three paradigms for reusing DNNs:
  conceptual reuse,
  adaptation reuse,
  and
  deployment reuse~\citep{davis2023JVA}. 
Prior work has characterized conceptual reuse in the form of DNN model reengineering and proposed the challenges in this reuse tpye, including performance debugging, and portability of deep learning operations~\citep{Jiang2023CVReengineering}. 
In the context of adapting PTMs in the application, there are two main challenges faced by software engineers:
  (1) technical adoption challenges,
  and
  (2) decision-making challenges such as model selection and evaluation~\citep{davis2023JVA}. 
For deployment reuse, Jajal \etal characterized failures of deep learning model converters which could compromise model quality~\citep{jajal2023ONNXFailureStudy}

Recent empirical research highlights the popularity of PTM registries among engineers. They appreciate these registries for their well-organized problem domains and user-friendly APIs, which are vital for downstream applications~\citep{Jiang2022PTMReuse, Jiang2022PTMSupplyChain,taraghi2024deep}. Studies by Jiang \etal and others have identified distinct differences between traditional software package reuse and PTM package reuse. These differences include varied decision-making processes, unique attributes that facilitate reuse, and specific risk factors relevant in PTM contexts. The impact of PTMs on software engineering practices has been a focal point of recent studies~\citep{castano2023exploring, kathikar2023assessingHFVulnerabilities}. 
Gong \etal have explored the usage contexts of PTM packages via an exploratory study from model hubs, but there is still a substantial gap in understanding the detailed reuse of these models~\citep{gong2023intendedUsageofPTM}.
Our dataset complements these findings by providing a detailed mapping between PTM packages and downstream GitHub repositories. This enables further, more insightful analysis of PTM reuse and adoption trends.

\subsection{Importance of Queryable PTM Metadata}
\label{sec:ImportanceofQueryableMetadata}
PTM metadata has been applied for several tasks.
In the realm of AI model management, the effective utilization of metadata plays a crucial role, such as helping with model auditing for assessing risks and ensuring responsible AI deployment~\citep{ raji2020closingtheAIAccountabilityGap}. 
Studies have shown that engineers often rely on various metadata types, such as evaluation metrics and hyperparameters, for informed model selection, underscoring their significance in the process.
Existing techniques effectively extract key metadata, supported by research papers, including model names, datasets, and frameworks~\citep{IBM2020AIMMX, Tsay2022AIMetadataExtractionIBM}. 
However, these methods do not support extraction from model cards, and not work for a comprehensive list of metadata (\eg hyperparameters, model size, hardware specification)~\citep{Montes2022DiscrepanciesAmongPTNN, castano2023analyzing}.
 The acquisition of extensive, queryable metadata types is crucial for enhancing model search, reuse, comparison, and composition~\citep{Li2022MetadataRepresentation4QueryableMLModelZoos}.
 
The evolving landscape of model repositories presents new challenges for metadata extraction~\citep{Jiang2022PTMReuse, Li2022MetadataRepresentation4QueryableMLModelZoos}. 
The main problem is the greater number of kinds of artifacts in this context, and linking them together with corresponding GitHub repositories is academic papers are hard.
Traditional methodologies have focused on model repositories on platforms like GitHub and academic papers~\citep{IBM2020AIMMX, Tsay2022AIMetadataExtractionIBM}.
Some platforms have tried to link papers to the relevant code repositories and models together, such as PapersWithCode~\citep{MetaAIpaperswithcodeabout}.
However, PTMs on model registries do not always link to GitHub projects and original research papers~\citep{Jiang2022PTMReuse}.
To address this gap in extracting metadata from model registries, we augment \DatasetNickname by leveraging state-of-the-art LLMs for metadata extraction. Capitalizing on the advanced capabilities of LLMs, we employ them to interpret and analyze model cards, effectively extracting pertinent metadata. 

\subsection{Open-Source PTM Datasets and Other Large-Scale Software Datasets} \label{sec:background-OOSDataset}

There are two existing PTM datasets: \textit{PTMTorrent}~\citep{jiang2023ptmtorrent} and \textit{HFCommunity}~\citep{ait2023hfcommunity}.
Both provide data included in the Hugging Face model registry, offering insights into PTMs.
However, both lack queryable PTM metadata and do not cover downstream applications.
These limitations reduce the range of mining questions that can be posed.
\DatasetNickname addresses both limitations by including additional content, \eg extracted metadata from model cards and links to downstream GitHub repositories.


There are also many large-scale open-source software datasets, such as 
  GHTorrent~\citep{Gousios2012GHTorrent},
  SOTorrent~\citep{Baltes2018SOTorrent},
  and
  TravisTorrent~\citep{Beller2017TravisTorrent}.
These datasets offer long-term data availability and help researchers avoid API rate limits~\citep{Gousios2012GHTorrent}.
These datasets have been instrumental in improving our understanding of software engineering,
  \eg of practices in
  continuous integration ~\citep{Beller2017ExplorativeStudyofTravisCI},
  static analysis~\citep{Zampetti2017HowOSProjectsUseStaticCodeAnalysisinCIPipelines},
  software development~\citep{baltes2019usage, cosentino2017systematic},
  and 
  testing~\citep{jiang2021CURE, Elsner2021RegressionTestOptimizationinCI}.

To advance our understanding of software engineering practices in deep learning systems, a large-scale, open-source dataset 
similar to those in previous studies 
is essential~\citep{Jiang2022PTMSupplyChain}. Such a dataset should encompass extensive software and its associated, queryable metadata for research purposes~\citep{Gousios2012GHTorrent, Baltes2018SOTorrent}.
Additionally, it should include dependency information to effectively characterize the software supply chain and keep the data updated. 
\DatasetNickname has a broader scope by including downstream GitHub applications, updated metadata, and recent models. Notably, our dataset incorporates a substantial number of large language models (LLMs) like Llama 2, which were absent in prior datasets.

\section{The PeatMOSS Dataset} 
\label{sec:original-Peatmoss}

This section summarizes and details the \DatasetNickname creation process.




 \begin{figure*}[t]
    \centering
    \includegraphics[width=1\textwidth]{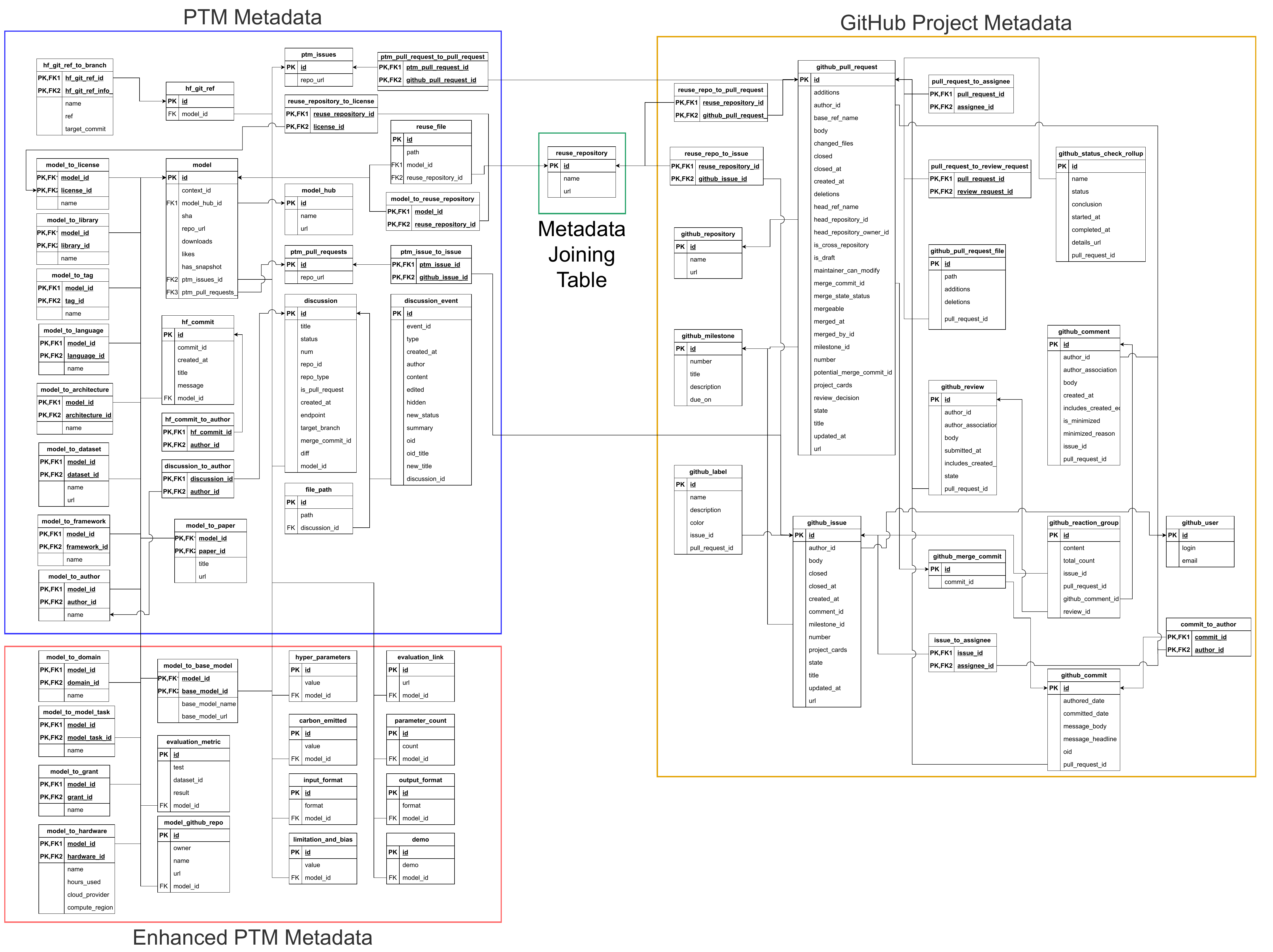}
    \caption{
    \small
    \DatasetNickname data schema. 
    There are four regions:
      tables for PTMs (basic~\cref{sec:original-Peatmoss} and enhanced~\cref{sec:enhanced-peatmoss}),
      tables for GitHub projects,
      and
      a table of PTM-Application dependency relations.
    Tables link to PTM and GitHub snapshots in a Globus share. 
    Our artifact has a navigable version (\cref{sec:DataAvailability}). 
    }
    \label{fig:DataSchema}
\end{figure*}

\subsection{Overview}
We created the \DatasetNickname dataset to enable study about \DatasetNicknameExpanded.
As illustrated by~\cref{fig:Overview},
\DatasetNickname comprises snapshots of PTMs and open-source repositories utilizing PTMs, as well as a mapping of PTMs to projects. 
For both PTMs and GitHub projects, \DatasetNickname contains metadata
(commits, issues, pull requests) and data (\eg model architecture and weights; \code{git} repositories), primarily collected in July-August 2023. 
\cref{fig:DataSchema} presents a uniform schema for retrieving PTM and project metadata is provided to facilitate analysis of PTMs and their use in open-source software projects.
Most information is indexed; some is stored as blobs.


\DatasetNickname contains the metadata of
  \TotalNumberOfPackagesMetadata PTM packages (\NumberOfHFPTMs from Hugging Face and \NumberOfPTPTMs from PyTorch Hub),
  \GitHubReuseRepoCountTP GitHub projects that use PTMs as dependencies,
  and \TotalNumberOfLinks links from these GitHub repositories to the PTMs they depend on. 
The dataset can be accessed in two formats. 
The ``metadata'' version of \DatasetNickname is a \NEWSizeOfMetadata SQLite database.
It contains the metadata of PTM packages and GitHub projects, and Globus links to their snapshots.
The \NEWDatasetSize ``full'' version has these snapshots:
    (1)
      the PTM package contents in each published version,
    and 
    (2)
      \code{git} history of the main branches of the GitHub projects.
\subsection{Dataset Creation Methodology} \label{sec:DatasetCreationMethodology}
Here we outline the methodology employed to compile \DatasetNickname,
  detailing PTM collection in~\cref{sec:PTMCollection},
  and
  the approach for associating PTMs with downstream GitHub repositories in~\cref{sec:DataCollection-Linking}.

\subsubsection{Collecting PTMs.} \label{sec:PTMCollection}

First, we must identify the model registries whose PTMs we will collect.
As discussed in \cref{sec:background-PTMSupplyChain}, there are three types of model registries.
Of these, only the open and gated types are open-source. 
For mining, we need registries that have APIs with recognizable signatures, allowing us to trace PTM-App dependencies (details in~\cref{sec:DataCollection-Linking}). 
Considering these criteria, we selected the most popular example from each open-source category that utilizes APIs. 
Thus, we included PTMs from Hugging Face (an open registry) and PyTorch Hub (a gated registry).
Hugging Face contains far more PTMs than PyTorch Hub, which influenced several decisions we made in creating \DatasetNickname.

Our PTM data collection includes three parts:
    (1) We saved \NEWNEWnumberOfPTMRepos
    PTM snapshots. This included the most popular PTM packages (\ie with over 50 downloads) on Hugging Face, and all PTMs on PyTorch Hub.
    This part of the data can provide a comprehensive view of PTM packages.
    (2) Among these ``full'' metadata, \TotalNumberOfLinks links
    from the PTMs to the downstream GitHub repositories have been identified. This part of the data can be connected to downstream GitHub data and allows miners to analyze the relationship between them.
    (3) For all PTMs hosted on Hugging Face and PyTorch Hub, we retrieved their metadata, resulting in a total number of \TotalNumberOfPackagesMetadata PTM package metadata being included in \DatasetNickname. 


\myparagraph{Soundness and Completeness:}
\DatasetNickname is comprehensive in terms of popular PTM packages, as it includes snapshots of those with over 10,000 downloads on Hugging Face. This provides a full view of widely-used PTMs and their connections to downstream GitHub projects, facilitating in-depth analysis. Additionally, the dataset includes metadata from all other PTMs on Hugging Face, which can be used for metadata-based analyses. 
\DatasetNickname enhances the diversity of PTM data by incorporating PTM packages from PyTorch Hub, including all available model repositories and their associated pull requests and issues.

\myparagraph{Implementation:} \label{sec:PTM_Implementation}
Metadata is collected using an \textit{Extract-Transform-Load (ETL)} pipeline for each model hub.
We first \textit{Extract} metadata from each model hub's API.
Then we \textit{Transform}, using this metadata to collect additional information (\eg following links to get packages backed by GitHub repositories).
Data that fits the shared schema is placed in an intermediate representation, while other data is preserved as a blob.
Results are \textit{Loaded} into our database.

\subsubsection{Collecting Downstream GitHub Repositories}\label{sec:DataCollection-Linking}
To enable research on \textit{PTM-PTM} and \textit{PTM-App} dependencies in open-source software projects, \DatasetNickname includes GitHub repositories that use at least one PTM from the two registries we captured.
We obtained the \GitHubReuseRepoCountTP pertinent GitHub repositories that existed as of \GitHubReuseRepoSourceGraphDate.
These repositories have an average 
of \GitHubReuseRepoSourceGraphAvgStar stars. 

For each of these \GitHubReuseRepoCountTP GitHub repositories, \DatasetNickname contains:
  (1) a full git clone;
  (2) all issues and associated metadata (as obtained through the GitHub CLI);
  and
  (3) all pull requests and associated metadata (via GitHub CLI).
We link them to the PTMs they use that were collected in~\cref{sec:PTMCollection}, to the extent possible with static analysis.

\begin{figure}
    \centering
      \fbox  {
    \includegraphics[width=0.98\linewidth]{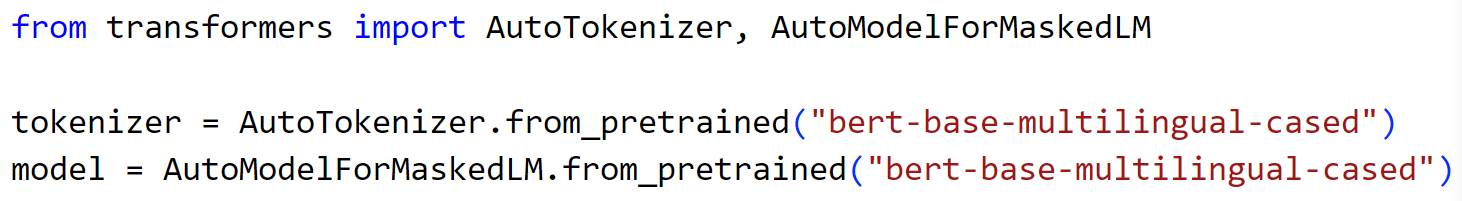}}
    \caption{
    Example use of two HuggingFace PTMs.
    The code initializes a tokenizer (\code{AutoTokenizer}) and a model (\code{AutoModelForMaskedLM}) from the \code{transformers} library for a multilingual BERT model.
    }
    \label{fig:usage_HF}
\end{figure}

The main challenge for this part of the dataset is identifying the GitHub repositories that use PTMs.
This task is non-trivial given the
  lack of standardized documentation or explicit labeling of PTM usage in repositories.
We devised an approach to automatically identify downstream GitHub repositories that depend on PTMs.
There are four steps to our approach:


\myparagraph{(Step 1) Signatures of PTM Use:}
The primary way to use PTMs from model hubs is through hub APIs.
There are many model hub libraries that access these APIs to retrieve PTMs by name.
\cref{fig:usage_HF} gives an example of accessing PTMs from Hugging Face via its \code{Transformers} library. 

We therefore define the \textit{signature} of PTM usage in an application as the combination of
  (1) library import and
  (2) calls into that library to load a PTM.
We specifically focus on signatures associated with Python libraries, as Python is the dominant language 
for PTM applications and almost all supported PTM loading libraries are written in Python~\citep{Tan2022DLSupplyChain,raschka2020MLinPython}.
We manually identified libraries and signatures in the documentation for the two target model hubs, Hugging Face~\citep{HFLibDoc}
and PyTorch~\citep{PytorchHub}.
In total, we found \GitHubReuseSignatureTotal signatures from \GitHubReuseLibrariesTotal Python libraries that access these hubs.

\myparagraph{(Step 2) Preliminary repository collection:}
We developed search patterns for each signature, and matched them against the content of files within GitHub repositories.
We searched for signatures in public, non-fork, non-archived repositories.
For this search, we used the \code{src} CLI tool from Sourcegraph, a popular code search engine that indexes GitHub repositories with $\geq5$ stars~\citep{SourcegraphDoc}. 
For example, a query for one of the signatures from the Diffusers library is: \code{``src select:file visibility:public count:all lang:Python content:`from diffusers' AND from\_pretrained(''}.

\begin{figure}[h]
    \centering
    \includegraphics[width=1\linewidth]{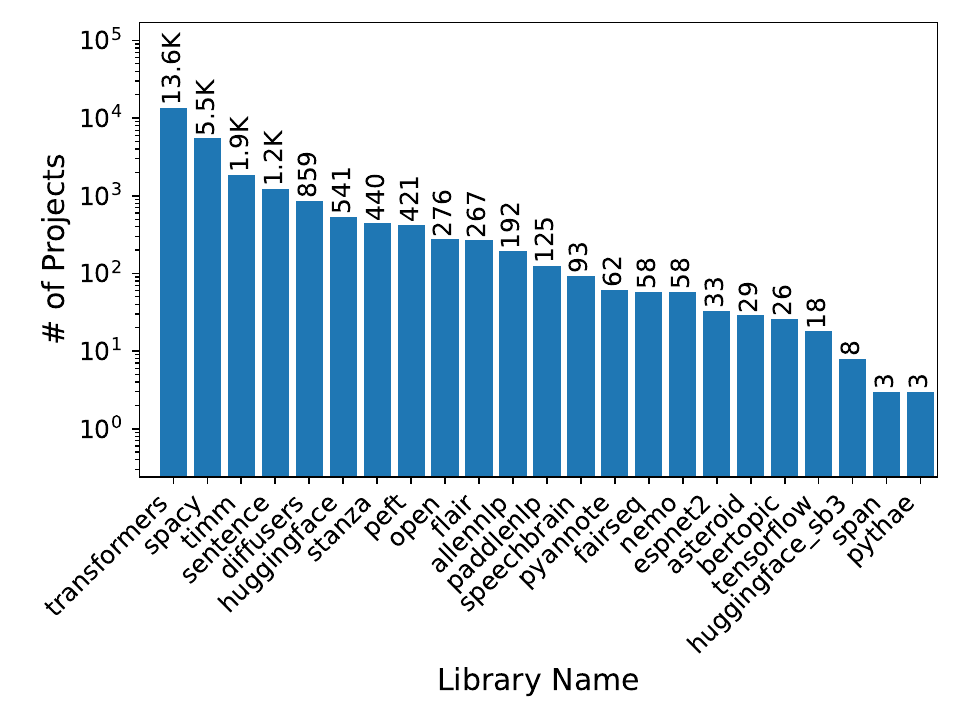}
    \caption{
    Number of projects that access PTMs from each Hugging Face library, as captured via Sourcegraph search.
    Note: Log scale.
    }
    \label{fig:NumberOfProjectsperLibrary}
\end{figure}

\myparagraph{(Step 3) Static Analysis:}
As Sourcegraph's search feature relies on text-based patterns, it is possible that some of the search results are false positives (\eg signatures that occur in commented-out code).
To mitigate this concern, we performed static analysis on the GitHub repositories from Step 2. 
This required some customization for each library.
Given the number of signatures (\GitHubReuseSignatureTotal signatures over \GitHubReuseLibrariesTotal libraries), we focused on the most popular libraries.
For PyTorch Hub, there are four libraries --- \code{torchvision}, \code{torchaudio}, \code{torchtext}, and direct uses of \code{torch} --- and we handle all associated signatures.
For Hugging Face, there are 23 libraries.
\cref{fig:NumberOfProjectsperLibrary} shows the distribution of usage:
we used signatures for the top five libraries 
  (\code{Transformers}, \code{SpaCy}, \code{Sentence-Transformers}, \code{Diffusers}, and \code{Timm}).
These accounted for 96\% of all downstream repositories that contain Hugging Face signatures according to our Sourcegraph search.

We performed static analysis using the Scalpel framework~\citep{li2022scalpel}. 
For each relevant source code file associated with a specific function signature, we construct an abstract syntax tree and extract the function calls contained within the file. 
Subsequently, we cross-reference the extracted functions with our predefined signatures which gives us a total of \GitHubReuseRepoCountTP repositories. 

\myparagraph{(Step 4) Mapping PTM-App relationship:}
Finally, we want to map which GitHub repositories depend on which PTMs.
For the function calls from each signature that load PTMs (identified in step 1), we extracted the function arguments (one of which is the PTM name), enabling us to extract specific PTMs being used in downstream GitHub repositories. 
We identify repositories that statically call the collected PTMs ---
\GitHubReuseRepoCountStaticMatch GitHub repositories do so, loading \NumberOfReusedPTMs distinct PTMs. 
Note that a PTM may be used by multiple repositories, and a repository can use multiple PTMs. 

\section{Enhanced PeaTMOSS via Metadata Extraction}
\label{sec:enhanced-peatmoss}

This section enhances \DatasetNickname by extracting indexed metadata from the unstructured metadata available in raw PTM packages.
As discussed in~\cref{sec:ImportanceofQueryableMetadata}, PTM metadata enables research and supports engineers' reuse process.
Past work observed that PTM metadata is often available in model cards, but unstructured, hampering ecosystem analysis~\citep{IBM2020AIMMX, Piorkowski2023QuantitativeAIRiskAssessment,Jiang2022PTMReuse}.
Our focus was on extracting metadata from Hugging Face PTM packages due to several reasons: 
    (1) a larger quantity of PTM packages, 
    (2) a larger quantity of mine-able documentation (model cards), and 
    (3) the centralized accessibility of their model cards for collection purposes.

We propose to use Large Language Models (LLMs) to extract metadata from model cards. 
Recent studies have demonstrated the versatility of LLMs in various tasks, including information retrieval~\citep{Bubeck2023AGISystem, gilardi_chatgpt_2023}. 
LLMs are effective in the task of metadata extraction from scientific documents~\citep{dunn2022structuredIEfromComplexSciTextwithFTLLM}.
In this work, we use ChatGPT, a leading commercial LLM~\citep{noauthor_chatgpt_nodate}. 

We identified desirable metadata through reviewing the literature and assessing available data in recent model cards, as shown in \cref{tab:MetadataMetrics}.
Prior works on metadata extraction indicate metadata of interest.
We supplemented those lists with metadata inspired by IBM's AI FactSheet~\citep{arnold2019AIfactsheets}, as well as observations from 50 recent model cards. These additional metadata include carbon emissions, model size, base model, limitation and biases, demonstration, grant/sponsorship information, and language.

\begin{table}[h]
\centering
\small

\caption{
  A list of PTM metadata mapped to the first paper that mentioned it. 
  \DatasetNickname includes these fields plus more (last row of table).
  Our artifact shows the mapping of these fields to our schema.
  }
\begin{tabular}{lp{0.6\linewidth}}

\toprule

\textbf{Paper}                        & \textbf{Newly Introduced Metadata of Interest} \\ 
\midrule
Schelter \etal, 2017~\citep{schelter2017automaticallyTrackingMetadataandProvenanceofMLExperiments}  & Model name, model version, framework, tags, dataset name, dataset version, dataset statistic, data transform, input/output format, evaluation, training time, environment, hyperparameters, prediction metadata          \\
Tsay \etal, 2020~\citep{IBM2020AIMMX}  & Reference, domain, has README, uses Python, popularity          \\

Li \etal, 2022~\citep{Li2022MetadataRepresentation4QueryableMLModelZoos}     & Model architecture, task, hardware       \\ 

Tsay \etal, 2022~\citep{Tsay2022AIMetadataExtractionIBM}    & Description, code, training job, training output, provenance          \\
\midrule

\DatasetNickname & Carbon emitted, model size, license, base model, limitation and biases, demonstration, grant/sponsorship, language (NLP)         \\ 
\bottomrule
\end{tabular}

\label{tab:MetadataMetrics}
\end{table}

\subsection{Prompt Design}
\label{sec:LLM4Metadata-PromptDesign}



Prompting provides the instructions to the LLM.
We followed the prompt design flow proposed by Zamfirescu \etal~\citep{zamfirescu2023johnny}, and outlined a structured approach for extracting and filling out a detailed metadata schema for models from Hugging Face.
To enhance our pipeline's performance, we use iterative prompting to test random sampled models~\citep{jha2023dehallucinating}. 
For metadata extracted with lower accuracy, we identified incorrect patterns, such as erroneous output formats and misleading results, to subsequently refine the corresponding prompts.
Moreover, we meticulously recorded instances where the model erroneously inferred metadata, known as hallucinations~\citep{sohail2023decodingChatGPT}, in the absence of relevant information. We also tracked cases where the model failed to extract information that was indeed present. Analyzing these outcomes enables us to pinpoint the metadata types that pose greater extraction challenges, thereby informing and refining our strategies in prompt engineering.
The prompts for two pipelines are all available in \cref{sec:DataAvailability}. 

\begin{figure*}[h]
    \centering
    \includegraphics[width=0.98\linewidth]{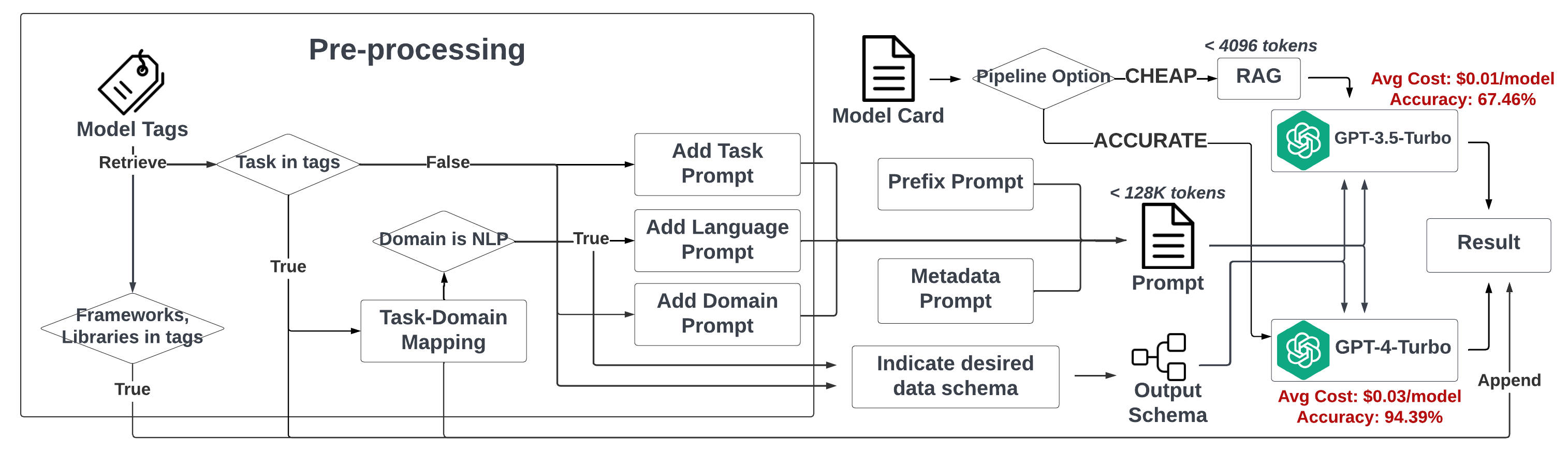}
    \caption{
     The ``cheap'' and ``accurate'' pipelines for metadata extraction.
     First the prompts are refined by the type of model.
     Then these prompts are applied to the model card.
     The cost-optimized ``cheap'' pipeline differs primarily by incorporating the Retrieval-Augmented Generation (RAG) framework~\citep{lewis2020RAG} to reduce token count. 
    }
    \label{fig:ExtractionPipeline}
\end{figure*}

The input prompt to our pipeline includes multiple components, as illustrated in \cref{fig:ExtractionPipeline}. 
The prefix prompt provides the domain and model background, setting extraction rules and schema adherence, with empty properties for absent document elements. 
The metadata prompt defines all the extraction requirements and formats for each metadata. 
The data schema is a formatted \code{json} file that used to store the extracted metadata.
If domain and tasks are not pre-processed from model tags, we include domain and task prompts, specifying domains (\eg multimodal) and tasks (\eg text-to-image).
For 
NLP models, language prompt is added to detail supported languages and extraction expectations (\eg Arabic, Chinese, Python).

\subsection{LLM Pipeline Design} \label{sec:LLMPipelineDesignNew}
We designed two LLM pipelines, one optimizing monetary cost and the other accuracy.
\cref{fig:ExtractionPipeline} summarizes these pipelines.

\myparagraph{Cheap Pipeline:}
By employing the Retrieval-Augmented Generation (RAG) strategy~\citep{lewis2020RAG} to mitigate the token usage for each model card and using the more cost-effective \code{GPT-3.5-turbo}, we have developed an efficient ``cheap'' pipeline. 
The \code{GPT-3.5-turbo}'s token limit of 4,096 tokens per request necessitates a method to extract complete metadata in segmented operations. 
The RAG strategy
helps the LLM incorporate relevant information from a knowledge base, providing contextual support and reducing the risk of generating inaccurate or speculative content. 
The RAG strategy reduces token usage, thus enhancing efficiency and reducing both computational and financial costs.

\myparagraph{Accurate Pipeline:}
The accurate pipeline, utilizing \code{GPT-4-turbo}, has a substantial improvement on addressing the token limit issue, along with enhanced performance in data extraction~\citep{Bubeck2023AGISystem}. An analysis of the token count across all model cards revealed that their lengths fell within the new token limit of \code{GPT-4-turbo} (128,000 tokens). Leveraging the advanced capabilities of GPT-4~\citep{openai2023gpt4}, we streamlined our pipeline by removing the RAG component. This modification allowed for a more holistic understanding of each model card, thereby improving metadata extraction efficiency.



\subsection{Evaluation} \label{sec:LLMEval}

\myparagraph{Sampling:}
Our initial evaluation required the selection of ground truth models, for which we analyzed the distribution of model tasks in the \DatasetNickname database. To achieve a representative sample, we employed stratified random sampling and sampled 50 models for evaluation. The models from different domains use different evaluation metrics so we want to cover most cases in our evaluation.
In this approach, each model task functioned as a separate stratum. The sample size for each task was aligned with its proportional representation in the database.
We focused on models that ranked among the top 100 in terms of downloads for each task, ensuring they were included in our database.
We then carefully examined the information of these models by checking their model cards and manually created the ground truth metadata for them.

\myparagraph{Accuracy:}
We selected accuracy as our primary metric for evaluating model performance, considering the context of manual assessment. This metric provides a straightforward and reliable method to evaluate the extraction. 
To calculate the overall accuracy, we tracked the frequency of successful metadata extractions matching our manual answer against the total number of extractions.

\myparagraph{Results:}
Comparing our results with manually labeled data, the GPT-3.5-turbo based pipeline achieved an accuracy rate of \GPTThreeEval. This evaluation was conducted on a random sample of 50 model cards from the \DatasetNickname dataset. Notably, the average cost for the \textit{cheap} pipeline was \code{\$0.01/model}. In a subsequent re-evaluation using the identical dataset, the \textit{accurate} pipeline exhibited a significant improvement in accuracy, reaching \GPTFourEval. The average cost for the GPT-4-turbo pipeline was slightly higher, at \code{\$0.03/model}.
We have not evaluated the specific factors that enhanced GPT-4's performance in this context.
However, its excellent performance led us to conclude the evaluation at this stage. 

\subsection{\DatasetNickname Enhancement}

We enhanced the \DatasetNickname dataset by incorporating metadata obtained from the ``accurate'' LLM pipeline, focusing on models that have over 50 downloads --- consistent with the model set for which we have collected snapshots.
The enhancement not only add the metadata to our dataset, but also successfully identifies \PTMPTMDepencencyCount \textit{PTM-PTM} dependencies within the supply chain, pinpointing upstream base models linked to each model as specified in their model cards.
Running the ``accurate'' pipeline to extract these enhanced metadata took \textasciitilde\$400 and \textasciitilde 40 hours.

\begin{figure}[h]
\centering
\includegraphics[width=0.98\linewidth]{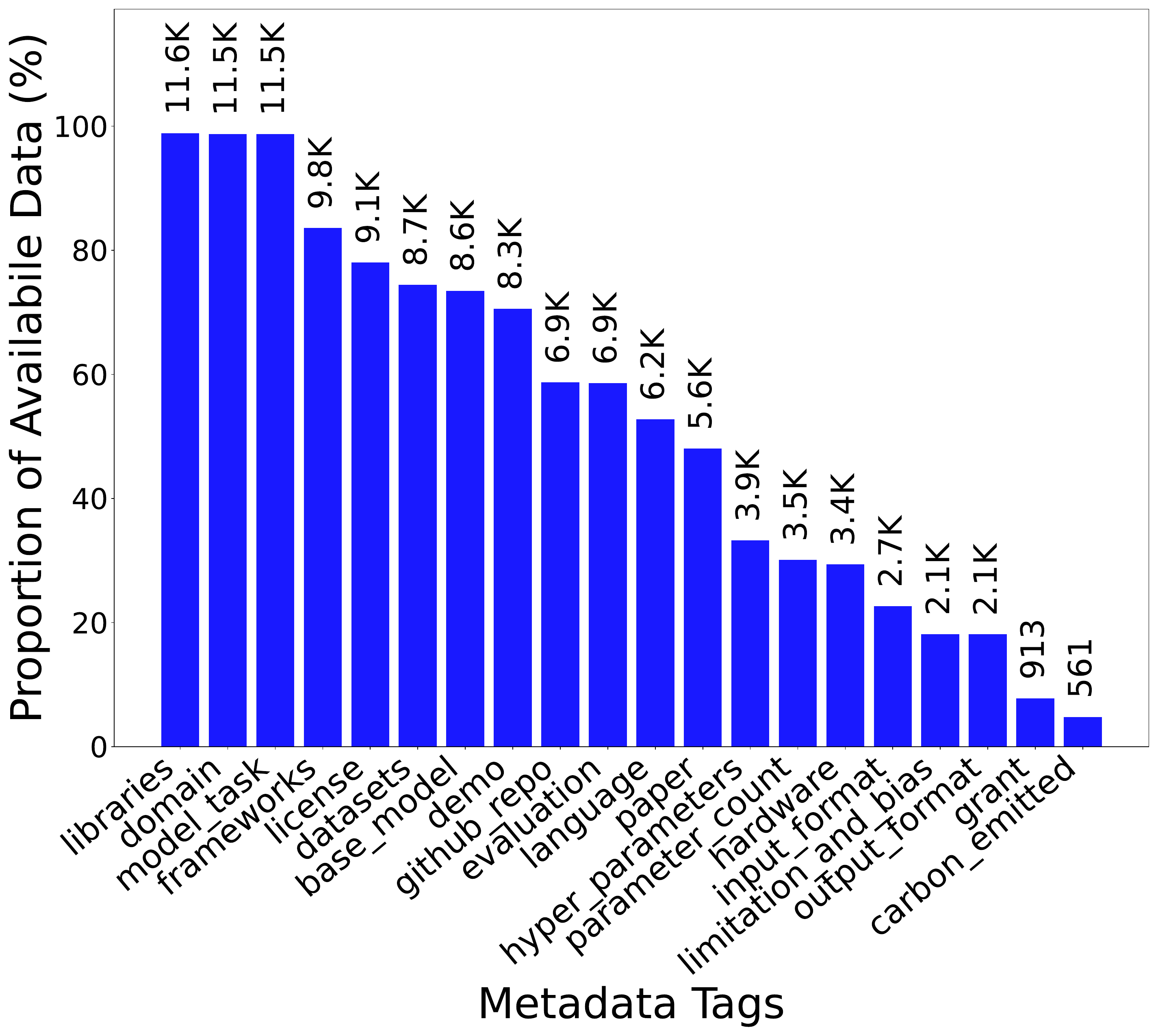}   
\caption{
  The proportion of available metadata for each category from Hugging Face model cards.
  Data is reported for the 11,975 models available at measurement time (Nov. 2023).
  About 17\% of the sample (2,321 models) became unavailable between project initiation and this measurement.
}
\label{fig:MetadataProportion}
\end{figure}

After metadata extraction, \cref{fig:MetadataProportion} shows the percentage of available metadata types for PTM packages.
Most models have metadata specifying libraries, domains, and model tasks, with 98.9\% for libraries and slightly less for domains and tasks. 
Metadata on frameworks, licenses, datasets, base models, demonstration, and evaluation are also prevalent, although to a lesser extent. 
On the other hand, less than half of the models include metadata on provenance (\ie github\_repo and papers). 
The data shows a significant absence of metadata concerning hyper-parameters, parameter count (\ie model size), hardware information, limitations, biases, and input/output formats, with these categories falling well below 40\%.
Less than 10\% model cards indicate the grant/sponsorship information and carbon emission.

\section{\DatasetNickname Initial Data Analysis} \label{sec:DataAnalysis}

We conduct some initial analysis of \DatasetNickname to illustrate its contents and measure the PTM supply chain. 
We report on the task domains of the PTMs (in aggregate and over time), PTM domains used by downstream GitHub repositories, and trends in model size.

\begin{figure}[h]
\centering
\includegraphics[width=0.95\linewidth]{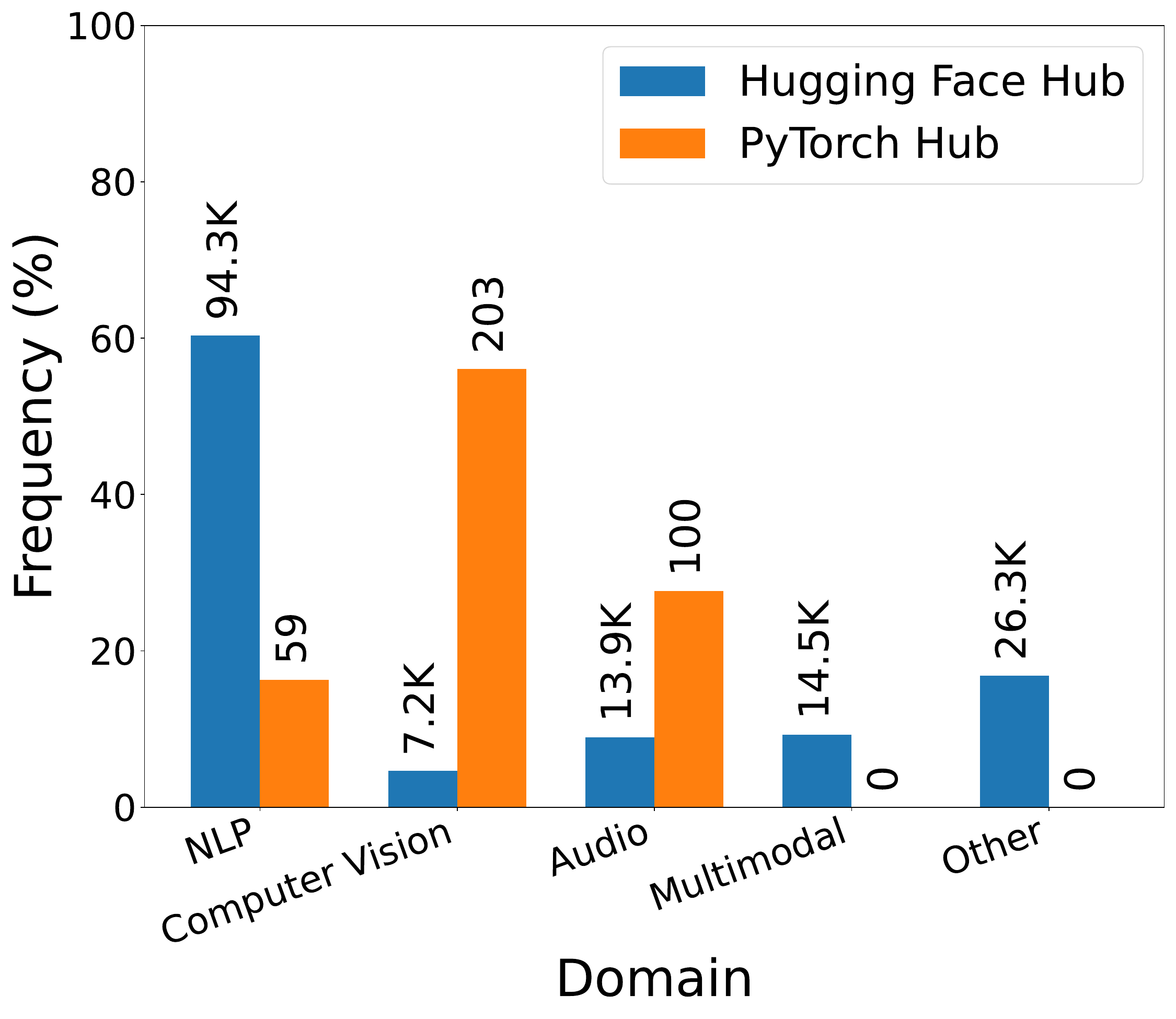}
\caption{
    The distribution and frequency of domains across PTMs in the two hubs. 
    For PyTorch Hub, we categorize labels such as research models, CUDA, and quantized models as ``Other'' for simplicity.
}
\label{fig:FreqofTaskDomain}
\end{figure}

\cref{fig:FreqofTaskDomain} presents the distribution of models across various problem domains in both Hugging Face and PyTorch Hub. It reveals that NLP models are predominant on Hugging Face (60.3\%), whereas PyTorch Hub features a higher frequency of CV (56.1\%) and Audio (27.6\%) models. 

\begin{figure}[h]
\centering
\includegraphics[width=0.95\linewidth]{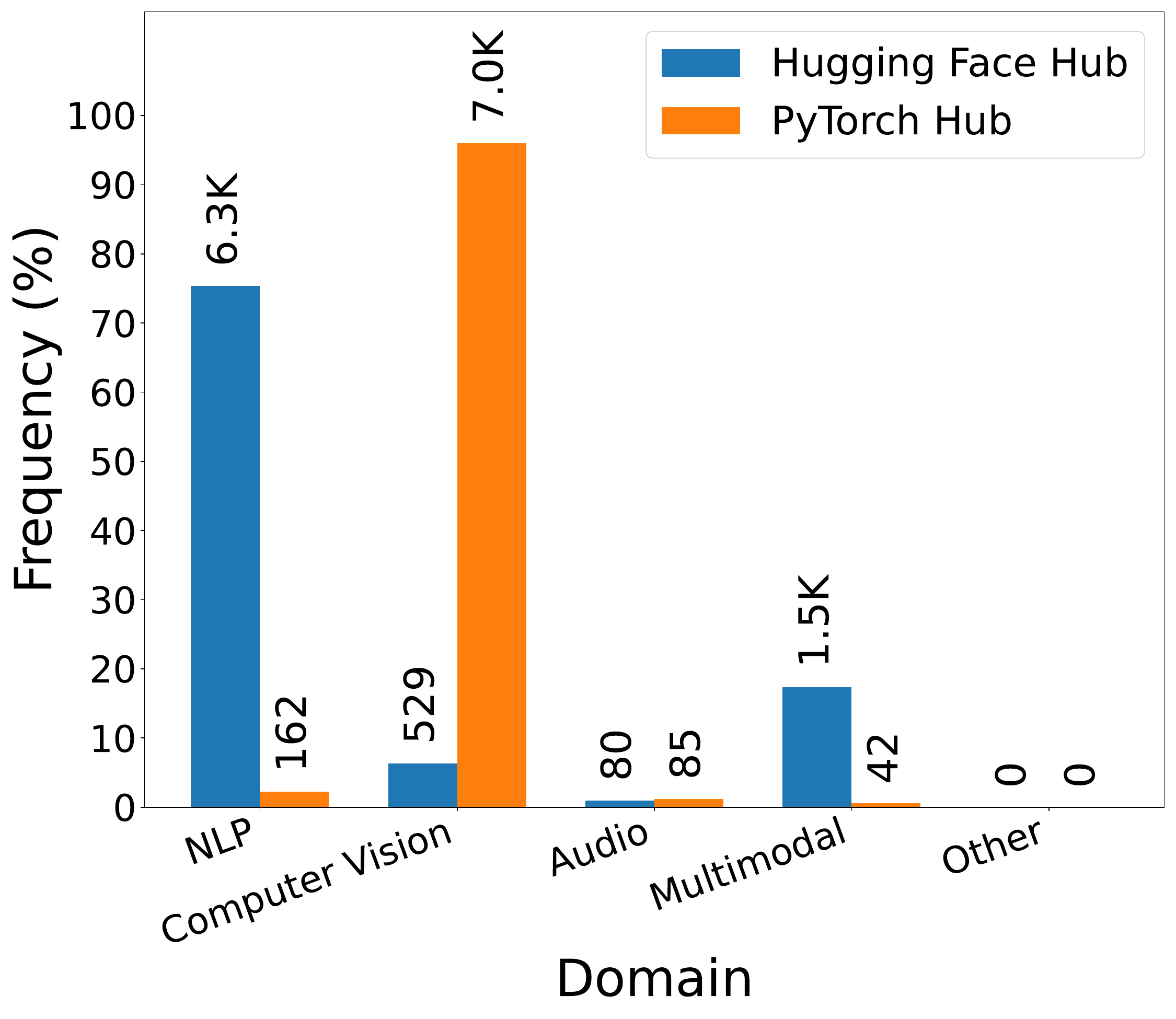}   
    \caption{
      Distribution and frequency of downstream GitHub repositories by model task.
      Hugging Face contains far more PTMs than PyTorch Hub, but they see comparable use in GitHub repositories.
    }
\label{fig:FreqofDownstreamApp}
\end{figure}

\cref{fig:FreqofDownstreamApp} displays the frequency of downstream GitHub repositories reusing PTM packages. It shows that NLP models are the most commonly reused on Hugging Face (75.4\%), followed by Multimodal (17.4\%) and CV (6.3\%) models. 
Conversely, PyTorch Hub users predominantly utilize CV (96.0\%), and only 2.23\% of them use NLP models.

\begin{figure}[h]
\centering
\includegraphics[width=0.95\linewidth]{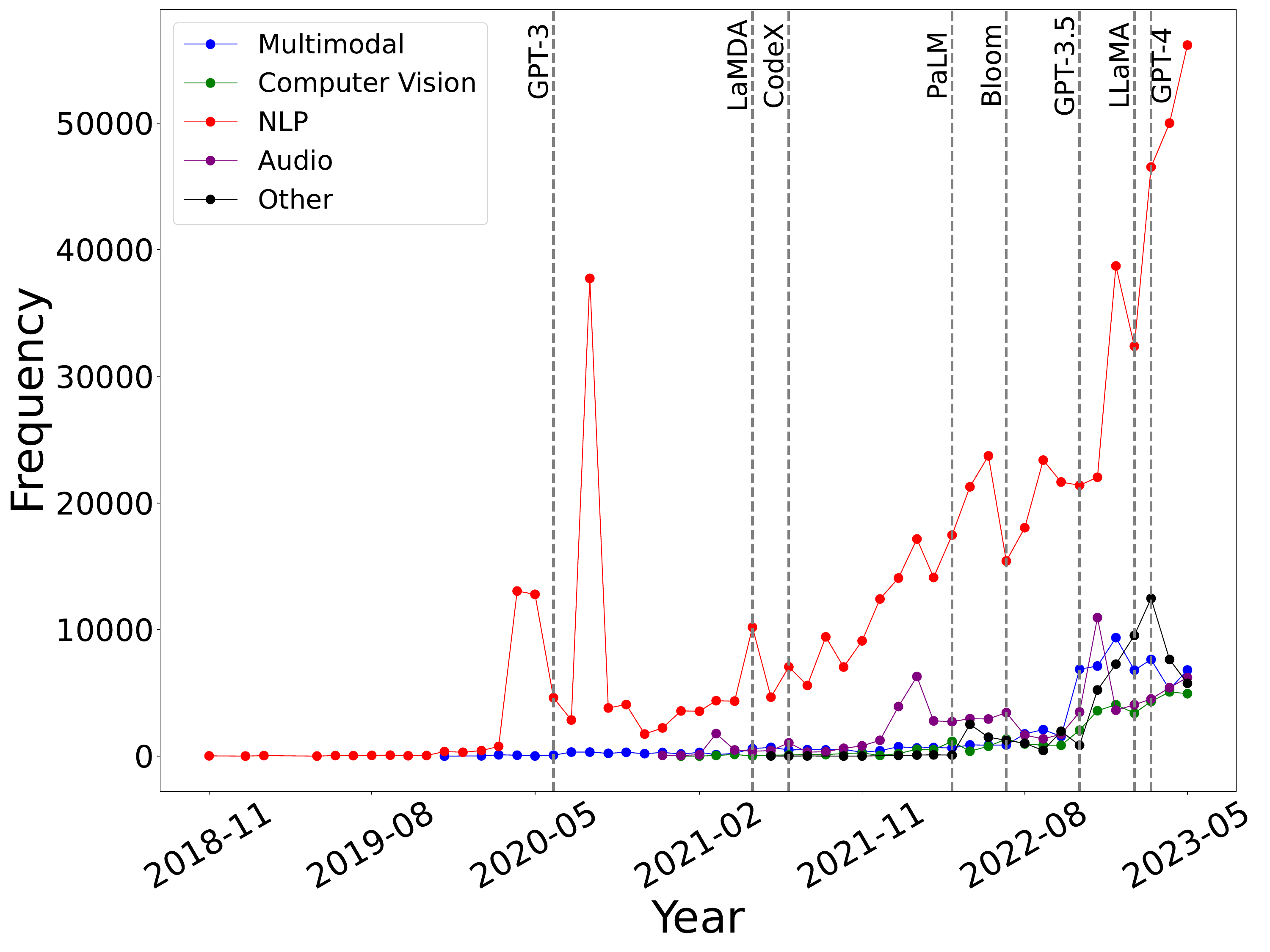}   
\caption{
The frequency of Hugging Face PTMs for different problem domains, tracked over time.
Vertical lines indicate events that may have caused the increase in parameters for NLP models.
}
\label{fig:TimeFrequency}
\end{figure}

\cref{fig:TimeFrequency} displays the creation frequency of Hugging Face PTM packages across various problem domains over time.
The data indicates a predominance of Natural Language Processing (NLP) models, likely reflecting Hugging Face's initial focus on NLP PTMs.
However, from August 2022 onward, packages for other domains have become more common. 
The jumps of NLP models in 2020 might relate to the rise of transformer family models during that time~\citep{nlplanet2021nlp_timeline}.

\begin{figure}[h]
\centering
\includegraphics[width=0.95\linewidth]{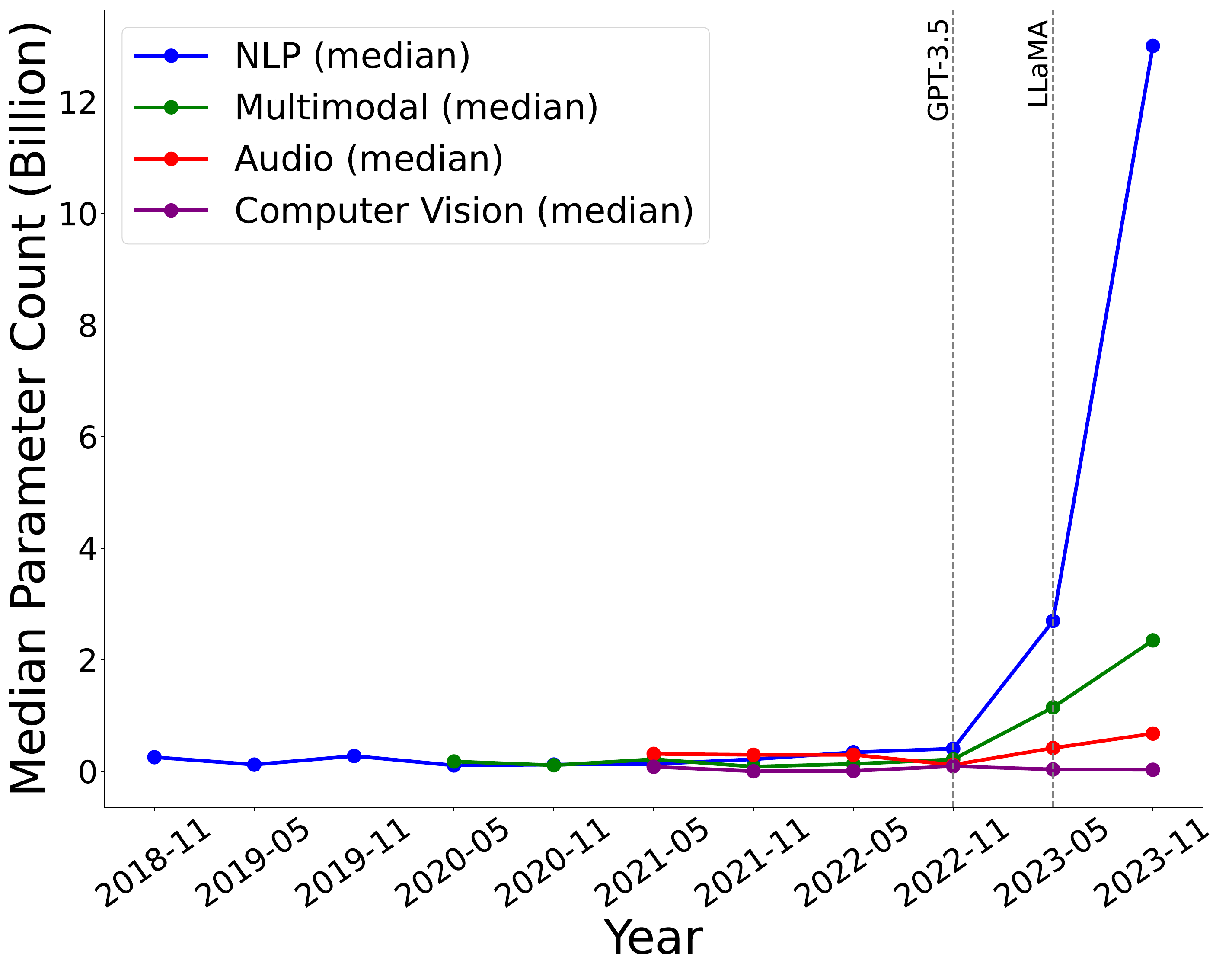}   
\caption{
Number of parameters (median) over time.
Vertical lines indicate landmarks of LLM models.
}
\label{fig:parameters}
\end{figure}

\cref{fig:parameters} tracks the median model size (\ie parameter count) by different domain.
There is a marked increase in the median size of NLP and multimodal PTMs, especially noticeable after March 2023.
Meanwhile, the median parameter count for Audio and CV models has remained relatively stable. 

\section{Mining PTM-App License Compatibility}
\label{sec:ExampleApplication}
This section illustrates the use of \DatasetNickname through a simple mining study on software licensing.
Software license information is important metadata for machine learning software~\citep{Bhatia2023ChangeTaxonomyforMLPipelines, Kuckertz2023MetadataBasedEcosystemtoImprovetheFAIRnessofResearchSoftware} and may promote responsible AI practices~\citep{contractor2022behavioral}.
When using a PTM, an engineer should comply with its license. 
We ask:
  \textit{(1) How do software licenses vary between PTMs and their downstream GitHub repositories?}
  and
  \textit{(2) How common are PTM-Application license incompatibilities?}


\subsection{Background on Software Package Licensing}


Licenses dictate the terms and conditions governing the reuse, modification, and redistribution of that software~\citep{laurent2004OSSLicensing}.
Licenses vary by the restrictions they place~\citep{GitHub_2022_Licenses}, \eg requiring derivative works to use a similar license (copyleft) or making the code freely available (public).
Integrating software with different licenses is complex~\citep{LicenseWiki} and may result in legal issues~\citep{rosen2005open,sojer2014understanding}. 
Studies of license incompatibility have been conducted in
  the Fedora Linux distribution~\citep{german2010understanding},
  Android applications~\citep{van2014tracing},
  and
  Java applications~\citep{german2012method,golubev2020study}),
as well as in
  multiple package ecosystems
    (\eg
      npm~\citep{qiu2021empirical},
      RubyGems~\citep{makari2022prevalence},
      and PyPI~\citep{xu2023understanding}).
We ask similar questions in the PTM ecosystem. 

We treat licensing definitions in PTMs comparable to other software packages~\citep{german2010understanding},
  with
  reuse (importing the PTM),
  modification (\eg fine-tuning a PTM),
  and
  redistribution (shipping the PTM in an application).
Following prior work, we treat mismatches as cases when there are different levels of license restrictiveness~\citep{wolter2023open}.

\subsection{License Measurement on \DatasetNickname}


\subsubsection{Method.}
In this analysis, we focus on the PTMs and GitHub projects in \DatasetNickname that are governed by a single license (7794, 54.5\%).
This model is simplistic~\citep{codescan} but aligns with GitHub's license API~\citep{GitHub_2022_Licenses}.
For PTMs, we use the license information from \DatasetNickname which was originally extracted using Hugging Face API from the model tags.
For downstream GitHub repositories, \DatasetNickname also includes license information that we extracted using the \code{codescan} tool in imitation of GitHub's \code{licensee} tool (\eg referencing files such as \code{LICENSE.txt}). 
For PTM-Application dependencies, we use the mapping given by \DatasetNickname.
We manually measured license compatibility 
based on the Linux 
Foundation's OSS license compatibility table~\citep{linuxfoundation2019fulfilling}.
%
In license pairings where no legal compatibility analysis was available, \eg in the case of ``no license''
(\PercentageOfAppNoLicense of the downstream GitHub repositories), 
we omit an assessment.

\subsubsection{Results.}



\begin{figure}[h]
\centering
\includegraphics[width=0.99\linewidth]{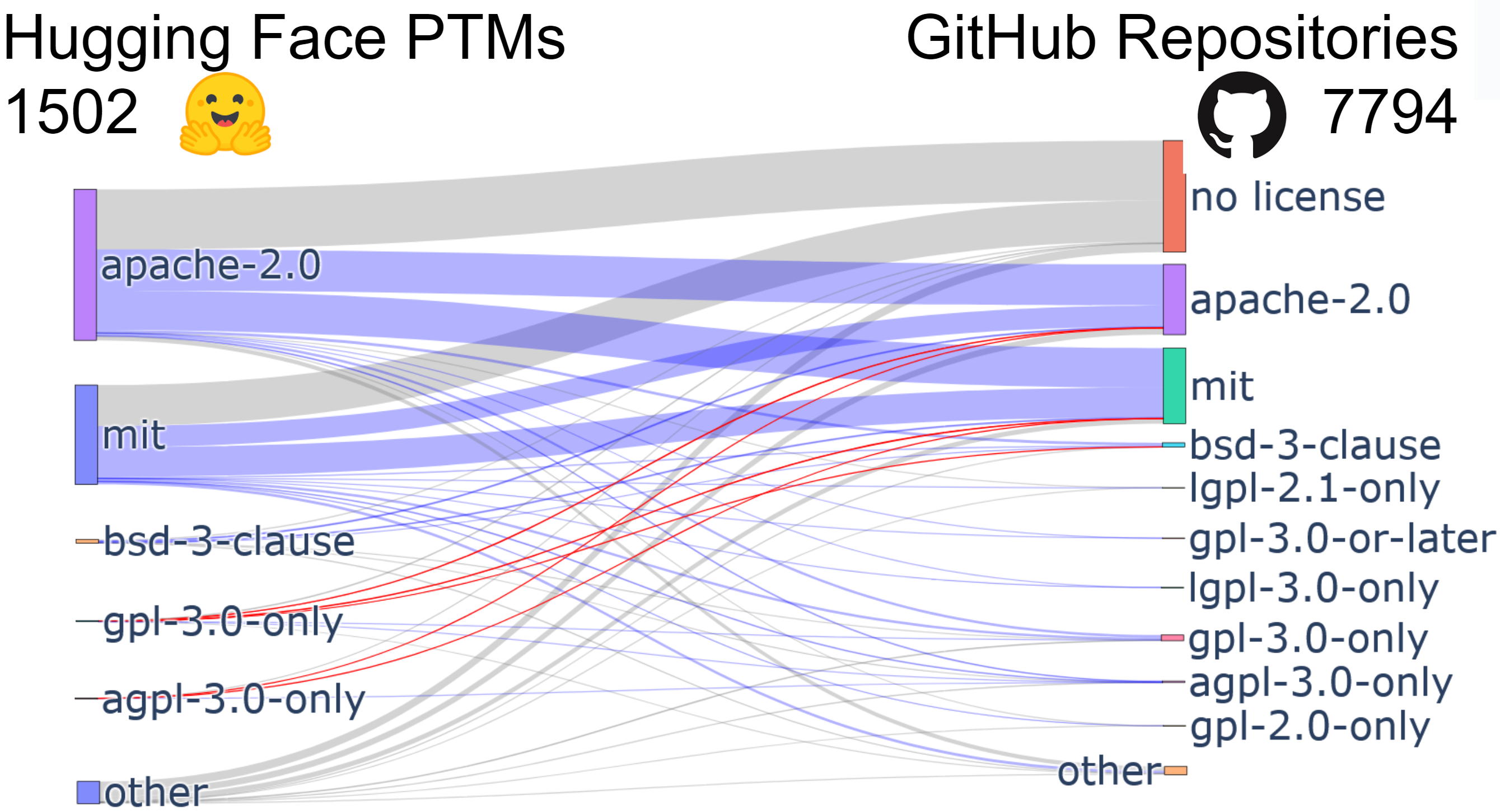}   
\caption{
Sankey Diagram for license compatibility.
Flows represent the licenses of PTMs and the downstream GitHub repositories that use them.
Blue flows are compatible, red are not.
Grey flows (\PercentageOfUnknownLicensePairs of pairs) represent license pairs that have not been analyzed by the Linux Foundation --- this is primarily caused by GitHub repositories lacking an explicit license. 
}
\label{fig:License}
\end{figure}

\cref{fig:License} answers our questions in a Sankey diagram, on the part of \DatasetNickname for which we have PTM-Application dependencies --- \NumberOfReusedPTMs PTMs used across \GitHubReuseRepoCount 
GitHub projects.

For license variation, we compare the left and right sides of~\cref{fig:License}.
The top-3 PTM licenses are Apache-2.0, MIT, and BSD-3-clause,
  while the top-3 GitHub repository licenses are MIT, Apache-2.0, and GPL-3.0-only.
Many downstream GitHub repositories (\PercentageOfAppNoLicense) choose not to define a license (``no license'' in the figure),
instead operating under the default posture of Hugging Face (full reuse~\citep{HF_license}) and GitHub (much stricter --- copyright reserved to author~\citep{GitHub_2022_Licenses}). 
In \PercentageOfIdenticalLicense of cases,
the PTM-Application licenses are identical.

For license compatibility,~\cref{fig:License} indicates compatible licenses with blue flows, otherwise red. 
In \PercentageOfInconsistentLicensePairs of PTM-Application dependencies, the licenses are incompatible.
In \DatasetNickname, this is 
  the result of copyleft provisions in the PTM's license that are not honored by the Application.

\section{Threats to Validity} \label{sec:threats}

Users of \DatasetNickname will inherit three types of threats to validity.

\myparagraph{Construct Validity.}
In the \DatasetNickname dataset, a key construct threat is the exclusion of conceptual reuse of PTMs~\citep{davis2023JVA}, which may limit our understanding of PTM usage. 
In our license analysis (\cref{sec:ExampleApplication}), 
  we assume that each project has one license, which aligns with GitHub's model (License API) but is imperfect.
Additionally, our metadata extraction method is supported by LLMs, which can pose threats on the reliability of our dataset. To mitigate this, we conducted an evaluation using stratified random sampling, observing accuracy of 94\% (\cref{sec:LLMEval}).

\myparagraph{Internal Validity.}
Internal validity threats in our study stem from the possibility of selection bias in curating PTMs and GitHub projects, as well as the changeable nature of repository contents over time.
Specifically, our LLM pipeline evaluation might be biased since it is based on a sample of only 50 models from Hugging Face. To mitigate this, we employed a stratified sampling technique. 
Additionally, when identifying the mapping from PTM to downstream GitHub repositories, our reliance on keyword searches on GitHub could miss other reuse signatures, potentially limiting the comprehensiveness of our findings.
To mitigate this, we employ distinct methodologies to manually gather the relevant signatures from Hugging Face and PyTorch Hub.

Another internal threat exists in our work due to the potential inaccuracies in the metadata extraction process from unstructured model cards. To mitigate this threat, we evaluated our extraction pipeline against a set of 50 manually-labeled model cards and found high accuracy. Moreover, we implemented a confidence threshold mechanism within our LLM extraction process. If the confidence level is below the set threshold, the data is earmarked for manual review, thereby improving the reliability of our dataset.

\myparagraph{External Validity.}
External validity threats are present due to the dataset primarily sourcing from Hugging Face and PyTorch Hub, which might not represent all PTM usage scenarios. 
To mitigate this threat, our dataset was designed to be expandable. 
We built our database in a flexible, modular way and provided clear, detailed documentation. This makes it straightforward to add new information to the database as needed. 
Another threat is present due to the selection of five libraries for the Hugging Face PTM downstream application collection. To mitigate the threat we check that 96\% of the primarily collected applications for Hugging Face PTMs belong to those five libraries. This fulfills the representativeness of our dataset.
Additionally, we acknowledge that there is another external threat on the dynamic nature of PTM data. The dataset will need to be updated --- we evaluated our data collection programs on multiple sites with comprehensive documentation provided for ongoing and future research.


\renewcommand{\arraystretch}{0.4}
{
\begin{table*}[th!]
\centering
\caption{
    Example lines of research for researchers to investigate, phrased as research questions.
    These questions are divided into three groups.
    The first group uses the Pre-Trained Model portion of the dataset (\emph{PTM}).
    The second group of questions makes use of the GitHub portion of the dataset (\emph{GH}).
    The third group asks questions that require Integrating both parts of the dataset (\emph{I}).
}
\label{table:RQs}
\begin{tabular}{p{15.2cm} c}
\toprule
\textbf{Research question} & \textbf{Related work} \\
\toprule


\\
\textbf{PTM-1:} What factors predict the popularity of a PTM? Intuition suggests that performance aspects such as accuracy and latency may dominate; what is the role played by factors such as software engineering quality? & \citep{borges2016understanding, lima2020characteristics, Jiang2022PTMReuse}
\\
\\

\textbf{PTM-2:} What naming conventions do PTMs follow? Are they consistent enough (within an architecture family? across families?) to support engineers looking for similar models?
& \citep{gresta2023naminginOOP, jiang2023PTMNaming}
\\
\\
\textbf{PTM-3:} PTM authors may reuse each others' work, \eg building off of model checkpoints or incorporating architectural building blocks. Is this forking, or a new form of software exchange? What is the phylogeny of the families of PTMs?
& \citep{Jiang2022PTMSupplyChain} \\
\\
\textbf{PTM-4:} There are many concerns about DNNs with unexpected or malicious behavior. How common are such DNNs? 
& \citep{Jiang2022PTMReuse, Wang2022Backdoor4TLwithPTM, Wang2022EvilModel2, guo2022threats}\\
\\
\textbf{PTM-5:} How to improve safety and security of model infrastructure, such as serialization formats and interoperability?
& \citep{jajal2023ONNXFailureStudy,david2024quack}\\
\\
\midrule
\\
{\textbf{GH-1:}} What kinds of defects are opened related to PTM use in the GitHub projects? How do these defects differ from defects opened on other aspects of the GitHub projects? & \citep{morovati2023buginMLSystems} \\
\\
{\textbf{GH-2:}} What do developers on GitHub discuss related to PTM use, \eg in the body text of issues and  
pull requests? What are developers' sentiments regarding PTM use?
Do the people issuing pull requests for PTMs have the right expertise?
 & \citep{yin2020team, sajadi2023interpersonalTrustinOSS}
 \\
 \\
\textbf{GH-3:} How often do developers change the PTM used to implement a feature? What factors influence this? & \citep{dilhara2021understanding}
\\
\\
{\textbf{GH-4:}} PTMs can underpin, enhance, or replace features implemented with traditional code. How common are these three modes of PTM adoption? How do PTMs subsequently affect the feature's failure modes? 
& \citep{li2021modeldiff, zhang2022ReMoS, qi2023reusingDNNthroughModelReengineering,nahar2023dataset} \\

\\
\midrule
\\

\textbf{I-1:} It can be difficult to interpret model popularity numbers by download rates. To what extent does a PTM's download rates correlate with the number of GitHub projects that rely on it, or the popularity of the GitHub projects? & \citep{fan2021makes} 
\\ 
\\
\textbf{I-2:} What are code smells for PTMs in the downstream GitHub repositories, and how do they affect these projects?
& \citep{zhang2022codeSmellsforMLApplications, van2021prevalence, cardozo2023prevalence}
\\
\\
\textbf{I-3:} What are application engineers' testing practices for their PTM-enabled features?
Do these vary based on the project's purpose, or the task delegated to the PTM?
How do testing practices acknowledge and address PTM stochasticity (\eg ``flakiness'')? 
& \citep{Li2022TestingMLSystemsinIndustry, Braiek2020onTestingMLPrograms,nahar2023dataset, eck2019understandingFlakyTest} \\
\\
\textbf{I-4:} How often do PTM application engineers update their PTM dependencies, \textit{e.g.,} due to (1) PTM deprecation, (2) PTM improvement, or (3) PTM antiquation (newer, better model)?
What is the typical technical lag for such updates? & \citep{ zerouali2018empiricalAnalysisofTechnicalLaginNPMDependencies, hora2018developers, wan2021MLCloudAPIs} 
\\
\\
\textbf{I-5:}
What are the characteristics of issue reports on PTM packages, \eg in terms of the kinds of questions asked, responsiveness of maintainers, issue density, and issue staleness?
How do these attributes differ from issue reports in GitHub repositories? & 
\citep{yang2023AIRepoIssues, Jiang2023CVReengineering} \\
\\
\textbf{I-6:}
What are the software signing requirements for PTMs? What are effective signatures for PTMs and training regimes?   & 
\citep{schorlemmer2024signing} \\

\bottomrule
\end{tabular}
\end{table*}
}

\section{Future Work} \label{sec:FutureWork}

Every dataset can be improved.
We highlight two enhancements for \DatasetNickname.
First, the current \textit{PTM-App} mapping relies solely on static analysis; a valuable extension would be to identify dynamic PTM usage.
Second, a deeper dive into the PTM supply chain would categorize the patterns of reuse in GitHub downstream repositories, such as direct loading versus fine-tuning or extending a model.
Adding these patterns would enrich the dataset.

\DatasetNickname enables many lines of research.
We highlight three lines in \cref{table:RQs}. 
The first line of research studies the Pre-Trained Model portion (\emph{PTM}). 
The second line of research studies the GitHub portion of the dataset (\emph{GH}). 
The third line integrates both parts (\emph{I}).

\section{Conclusion} \label{sec:conclusion}
Pre-Trained Models (PTMs) offer state-of-the-art performance in various domains, and are being incorporated into many computing systems.
PTMs represent a new frontier for mining software repositories, but the community lacks a comprehensive dataset.
To enable PTM mining, this paper presents the \DatasetNickname dataset, a collection of PTM metadata, PTM snapshots, downstream GitHub repositories that use PTMs, and mappings between PTMs and the repositories that use them.
To augment the data available from PTM registries and GitHub APIs, we developed an automated process to extract and standardize PTM metadata, enhancing the dataset's utility.
To demonstrate applications of \DatasetNickname, we present the first detailed statistics of the PTM supply chain, and examine software license inconsistencies between PTMs and their dependent projects.
For future work, we propose thirteen distinct research questions along three lines of research:
  studies of PTMs,
  studies of downstream use on GitHub,
  and
  studies that integrate data on PTMs and their dependents.

\section{Data Availability} \label{sec:DataAvailability}

The source code associated with this research is available at {\url{https://github.com/PurdueDualityLab/PeaTMOSS-Artifact}}, along with a demo version of the dataset.
The full \DatasetNickname dataset is stored on our organization's archival-grade storage system and accessible through Globus at {\url{https://transfer.rcac.purdue.edu/file-manager?origin\_id=ff978999-16c2-4b50-ac7a-947ffdc3eb1d\&jorigin\_path=\%2F}}.



\ifANONYMOUS
\else
\section*{Acknowledgments}
    This work was supported by
      gifts from Google and Cisco;
      NSF awards \#2107230, \#2229703, \#2107020, and \#2104319;
      and
      by the Faculty Research Participation Program at Argonne National Laboratory.
    We acknowledge the support of the Natural Sciences and Engineering Research Council of Canada (NSERC), RGPIN-2019-05071.
    This research used resources of the Argonne Leadership Computing Facility, a U.S. Department of Energy (DOE) Office of Science user facility at Argonne National Laboratory and is based on research supported by the U.S. DOE Office of Science-Advanced Scientific Computing Research Program, under Contract No. DE-AC02-06CH11357.
    We thank Purdue's Rosen Center for Advanced Computing (RCAC) for ongoing support in hosting the dataset.
    We thank A. Raghav, A. Qi, and Y. Mehta for their assistance, and members of the Purdue Duality Lab for their feedback on the manuscript.

\fi

\ifARXIV
\else
\raggedbottom
\pagebreak
\balance
\fi

\pagebreak

\bibliographystyle{ACM-Reference-Format}
\balance
\bibliography{bibliography/master}

\end{document}